\documentstyle[twoside,fleqn,npb,epsfig]{article}
\newcommand{\Eq}[1]{Eq.(\ref{#1})} 
\newcommand{\Fn}[2]{F_{#1}^{(#2)}(m_i^2,p^2)} 
\newcommand{\Pol}[2]{P_{#1,#2}(m_i^2,p^2)} 
\newcommand{\Qol}[2]{Q_{#1,#2}(m_i^2,p^2)} 
\newcommand{\D}{D(m_i^2,p^2)} 
\newcommand{\F}[1]{F_{#1}(n,m_i^2,p^2)} 
\newcommand{\dnk}[1]{ \frac{d^nk_{#1}}{(2\pi)^{n-2}} }

\newcommand{\AmS}{{\protect\the\textfont2
  A\kern-.1667em\lower.5ex\hbox{M}\kern-.125emS}}

\title{Numerical evaluation of master integrals from differential equations
\thanks{Supported in part by the EC network EURIDICE, 
contract HPRN-CT2002-00311.}
}

\author{M. Caffo\address{INFN, Sezione di Bologna, \\
                               I-40126 Bologna, Italy}
 \kern-2pt\address{Dipartimento di Fisica, Universit\`a di Bologna, \\
                               I-40126 Bologna, Italy}
\thanks{Speaker at RADCOR 2002 \& Loops and Legs in Quantum Field Theory, 
8-13 September 2002, Kloster Banz, Germany.},
        H. Czy{\.z}\address{Institute of Physics, University of Silesia, \\
        PL-40007 Katowice, Poland}
        and
        E. Remiddi$^{\rm b\ a}$}

\begin{document}

\begin{abstract}
The 4-th order Runge-Kutta method in the complex plane is proposed 
for numerically advancing the solutions of a system of first order 
differential equations in one external invariant satisfied by the 
master integrals related to a Feynman graph.
The particular case of the general massive 2-loop sunrise self-mass 
diagram is analyzed.
The method offers a reliable and robust approach to the direct and precise 
numerical evaluation of master integrals.
\end{abstract}

\maketitle

\section{Introduction}

The very high precision of present and planned particle physics experiments 
requires comparable or better accuracy on the theoretical side. 
This fact promotes developments of new methods in the calculations 
of radiative corrections, which are today a living and expanding field.

The nowadays widespread organization of the calculations is based on the 
integration by part identities and on the evaluation of the 
master integrals (MI) \cite{TkaChet}.
We believe that the systematic use of the differential equations for 
the MI, or Master Differential Equations (MDE), can be a viable method 
for their analytic calculations in many cases. 
In these cases, but also when the number of variables and parameters 
prevents the success of an analytic calculation, the MDE can still be 
profitably used for direct numerical evaluation of the MI. 
This is an alternative to the more commonly used integration methods 
or to the more recently introduced difference equations method. 

A method which uses the MDE to get a numerical solution, 
starting from a known value, is presented here and its features are 
discussed.

\section{Master Differential Equations}

Starting from the integral representation of the $N_{MI}$ MI, 
related to a certain Feynman graph, 
by derivation with respect to one of the internal masses $m_i$ \cite{Kotikov} 
or one of the external invariants $s_e$ \cite{Remiddi} and with the repeated 
use of the integration by part identities, a system of $N_{MI}$ independent 
first order partial MDE is obtained for the $N_{MI}$ MI. 
For any of the $s_e$, say $s_j$, the equations have in general the form 
\begin{eqnarray} 
&&K_k(m_i^2,s_e)
\ {\frac{\partial}{\partial s_j}} F_k(n,m_i^2,s_e) = \nonumber\\
&&{\kern-20pt}\sum_{l} M_{k,l}(n,m_i^2,s_e) F_l(n,m_i^2,s_e)
+T_k(n,m_i^2,s_e), \nonumber\\ 
&& k,l=1,...,N_{MI} \label{mde}
 \end{eqnarray} 
where $F_k(n,m_i^2,s_e)$ are the MI,  
$K_k(m_i^2,s_e)$ and $M_{k,l}(n,m_i^2,s_e)$ are polynomials, 
while $T_k(n,m_i^2,s_e)$ are polynomials times simpler MI of the 
subgraphs of the considered graph.
The roots of the equations 
\begin{equation} 
K_k(m_i^2,s_e) = 0 
\label{sp}
\end{equation} 
identify the {\em special} points, where numerical calculations are 
troublesome.  
Fortunately analytic calculations at those points come out to be 
possible in all the attempted cases so far.
They might not be simple and often require some external knowledge, 
like the assumption of regularity of the solution at that {\em special} point. 

To solve the system of equations it is necessary to know the MI for a 
chosen value of the differential variable, $s_j$ in \Eq{mde}.
For that purpose we use the analytic expressions at the {\em special} 
points, taken as the starting points of the advancing solution path. 
Moreover starting from one {\em special} point, not only the values of 
the MI are necessary, but also their first order derivatives at that point. 
That is because some of the coefficients $K_k(m_i^2,s_e)$ of the MI 
derivatives in the differential equations \Eq{mde} vanish at that point. 
Therefore also the analytic expressions for the first derivatives of MI
at {\em special} points are obtained, but this usually comes out to be a 
simpler task (unless poles in the limit of the number of dimensions $n$ 
going to 4 are present). 

Enlarging the number of loops and legs increases the number of parameters, 
MI and equations, but does not change or spoil the method. 

\section{The 4-th order Runge-Kutta method}

Many methods are available for obtaining the numerical solutions of 
a first-order differential equation \cite{RK} 
\begin{equation} 
\frac{\partial y(x)}{\partial x} = f(x,y) \ .
\label{fode}
\end{equation}
The Euler method advances the solution from a point $x_n$, where the solution 
$y_n$ is known, to the point $x_{n+1} = x_n + \Delta$ 
\begin{equation} 
y_{n+1} = y_n + \Delta f(x_n,y_n) +{\cal O}(\Delta^2)  
\label{euler}
\end{equation} 
omitting terms of order $\Delta^2$.
A direct improvement of the Euler method is the 4-th order Runge-Kutta method, 
that we choose, because it is considered a rather precise and robust approach. 
By suitably choosing the intermediate points where calculating $f(x,y)$ 
one obtains the 4-th order Runge-Kutta formula 
\begin{eqnarray} 
 k_1 &=& \Delta f(x_n,y_n), \ \nonumber \\
 k_2 &=& \Delta f(x_n+\frac{\Delta}{2},y_n+\frac{k_1}{2}), \ \nonumber \\
 k_3 &=& \Delta f(x_n+\frac{\Delta}{2},y_n+\frac{k_2}{2}), \ \nonumber \\
 k_4 &=& \Delta f(x_n+\Delta,y_n+k_3), \ \nonumber \\
 y_{n+1} &=& y_n +\frac{k_1}{6}+\frac{k_2}{3}+\frac{k_3}{3}+\frac{k_4}{6} 
+{\cal O}(\Delta^5) 
\label{4ork}
\end{eqnarray} 
which omits terms of order $\Delta^5$.

To avoid numerical problems due to the presence of {\em special} points 
on the real axis, it is convenient to choose a path for advancing the 
solution in the complex plane of $x$.

The extension from one first-order differential equation to a system of 
$N_{MI}$ first-order MDE for the $N_{MI}$ MI is straightforward \cite{RK}. 

\section{Results: sunrise, ...}

\begin{figure}[htb]
\vspace{9pt}
{\scalebox{.7}[.7]{\includegraphics*[80,20][370,120]{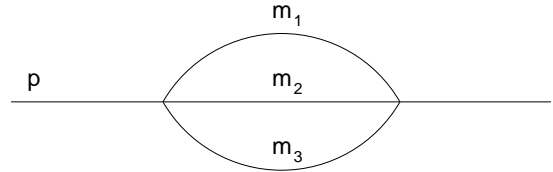}}}
\caption{The general massive 2-loop sunrise self-mass diagram.}
\label{fig:sunrise}
\end{figure}

To test the method we have chosen to start from the simple, but not trivial, 
2-loop sunrise graph with arbitrary masses \cite{CCLR1,CCR3}, shown in 
Fig.\ref{fig:sunrise}. 

\begin{figure}[htb]
{\scalebox{.7}[.7]{\includegraphics*[20,40][260,140]{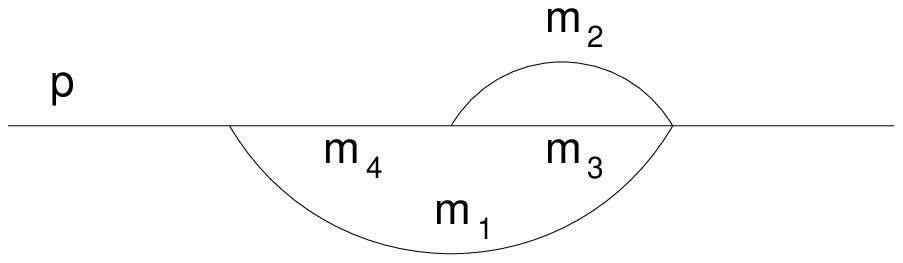}}}
\caption{The general massive 2-loop 4-denominators self-mass diagram.}
\label{fig:4den}
\end{figure}

\begin{figure}[htb]
{\scalebox{.7}[.7]{\includegraphics*[20,20][260,120]{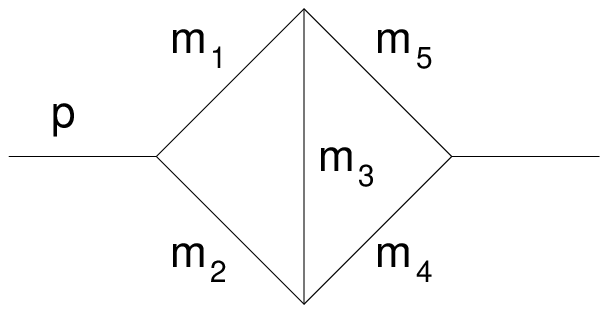}}}
\caption{The general massive 2-loop 5-denominators self-mass diagram.}
\label{fig:5den}
\end{figure}

This graph is one of the topologies of the 2-loop self-mass and has 
4 MI. The other topologies with 4 and 5 propagators, 
shown in Fig.\ref{fig:4den} and in Fig.\ref{fig:5den} respectively, 
have one more MI each \cite{Tarasov,CCLR1,CCLR2}. 
The calculations for these graphs are in progress \cite{CCGR}. 

The sunrise general amplitude can be written in the integral form as 
\begin{eqnarray} 
 &&{\kern-40pt}A(n,m_i^2,p^2,-\alpha_1,-\alpha_2,-\alpha_3,\beta_1,\beta_2) 
     = \nonumber \\  
 \mu^{8-2n} &&{\kern-30pt} \int \dnk{1} \int \dnk{2} \; 
      \frac{ (p\cdot k_1)^{\beta_1} (p\cdot k_2)^{\beta_2} } 
           { D_1^{\alpha_1} D_2^{\alpha_2} D_3^{\alpha_3} }, \nonumber \\
 D_1&=&(k_1^2+m_1^2), \nonumber \\
 D_2&=&(k_2^2+m_2^2),  \nonumber \\
 D_3&=&( (p-k_1-k_2)^2+m_3^2 ), 
\end{eqnarray} 
where $\mu$ is the arbitrary mass scale accounting for the continuous value 
of the dimensions $n$. As the natural scale of the problem is the 
threshold of the sunrise amplitudes, we take $\mu = m_1 + m_2 +m_3$. 

We choose for the 4 MI the amplitudes
\begin{eqnarray} 
  \F{0}{\kern-7pt}&=&{\kern-10pt}A(n,m_i^2,p^2,-1,-1,-1,0,0) \ ,\nonumber \\ 
  \F{1}{\kern-7pt}&=&{\kern-10pt}A(n,m_i^2,p^2,-2,-1,-1,0,0) \ ,\nonumber \\ 
  \F{2}{\kern-7pt}&=&{\kern-10pt}A(n,m_i^2,p^2,-1,-2,-1,0,0) \ ,\nonumber \\ 
  \F{3}{\kern-7pt}&=&{\kern-10pt}A(n,m_i^2,p^2,-1,-1,-2,0,0) \ , 
\end{eqnarray} 
which are connected by the relations $(j,i = 1,2,3)$
\begin{equation} 
  \F{j} = - \frac{\partial}{\partial m_j^2} \F{0} \ .  
\end{equation} 
The above relations are very important for analytic calculations, 
but are not used in the present numerical approach.

To obtain the system of differential equations for the 4 MI in the form of 
\Eq{mde} it is necessary to develop explicitly the integration by parts 
identities for the amplitudes with the values of the exponents $\alpha_i$ 
and $\beta_j$ satisfying the relations $ \sum_{i=1,2,3} (\alpha_i-1)=2 $
and $ \sum_{j=1,2} \beta_j=2 $. 

In the differential equations of the sunrise MI the only lower order diagram 
entering is the 1-loop vacuum graph
\begin{eqnarray}
 T(n,m^2) &=& \int \dnk{} \frac{1}{k^2+m^2} \nonumber \\
          &=& \frac{m^{n-2}C(n)}{(n-2)(n-4)}\ . 
\end{eqnarray}
The function
\begin{equation}
 C(n) = \left(2 \sqrt{\pi} \right)^{(4-n)} \Gamma\left(3-\frac{n}{2}\right)\ , 
\end{equation}
which appears in the expressions for the MI as an overall factor with an 
exponent equal to the number of loops, is usually 
kept unexpanded in the limit $n \to 4$, and only at the very end of the 
calculation for finite quantities is set $C(4) = 1$. 

When the sunrise MI are expanded in $(n-4)$, for $j=0,1,2,3,$ and 
$i=1,2,3,$
\begin{eqnarray} 
\F{j} = C^2(n) &&{\kern-25pt}\Biggl\{ \frac{1}{(n-4)^2} \Fn{j}{-2} \nonumber \\
 + \frac{1}{(n-4)}   \Fn{j}{-1} &+& \Fn{j}{0} \nonumber \\
 + {\cal O} (n-4) && \Bigr\} \ , 
\end{eqnarray} 
the coefficients of the poles can be easily obtained analytically for 
arbitrary values of the external squared momentum $p^2$, 
\begin{eqnarray} 
 \Fn{0}{-2} &=& -\frac{1}{8} (m_1^2+m_2^2+m_3^2) \ ,
\nonumber \\
 \Fn{0}{-1} &=&  \frac{1}{8} \Biggl\{ \frac{p^2}{4} 
                +\frac{3}{2} (m_1^2+m_2^2+m_3^2) \nonumber \\  
&& {\kern-90pt}
 - \left[ m_1^2 \log\left(\frac{m_1^2}{\mu^2}\right) 
 +m_2^2 \log\left(\frac{m_2^2}{\mu^2}\right)
 +m_3^2 \log\left(\frac{m_3^2}{\mu^2}\right) \right] 
                        \Biggr\} \ , \nonumber \\
 \Fn{k}{-2} &=& \frac{1}{8} \ , \quad k=1,2,3 \nonumber \\
 \Fn{k}{-1} &=&  - \frac{1}{16} 
 + \frac{1}{8} \log\left(\frac{m_k^2}{\mu^2}\right) \ . 
\end{eqnarray} 
The finite parts satisfy the differential equations
\begin{eqnarray} 
 p^2 \frac{\partial}{\partial p^2} \Fn{0}{0}{\kern-10pt}&=&
{\kern-10pt}\Fn{0}{0} + \Fn{0}{-1} \nonumber \\ 
  && {\kern-100pt} + \sum_{j=1,2,3} m_j^2 \Fn{j}{0} \ , 
\label{mde1}
\end{eqnarray} 
and ($i,j,k,l=1,2,3$, with $j \ne k \ne l$) 
\begin{eqnarray} 
 &&{\kern-20pt} 8 \D p^2 \frac{\partial}{\partial p^2} \Fn{l}{0} =  
 \nonumber \\ 
 && \phantom{+}  4   \D \Fn{l}{-1} \nonumber \\ 
 &&{\kern-20pt}+ \Pol{l}{0} \left[ 16 \Fn{0}{0} + 28 \Fn{0}{-1} 
\right. \nonumber \\ 
 && {\kern+100pt} \left. + 12 \Fn{0}{-2} \right] \nonumber \\ 
 &&{\kern-20pt}+8\Pol{l}{l} \left[ \Fn{l}{0} +\Fn{l}{-1} \right] \nonumber \\
 &&{\kern-20pt}+8 \Pol{l}{j} \left[ \Fn{j}{0} +\Fn{j}{-1} \right] \nonumber \\ 
 &&{\kern-20pt}+8 \Pol{l}{k} \left[ \Fn{k}{0} +\Fn{k}{-1} \right] \nonumber \\ 
 &&{\kern-20pt}+ \Qol{l}{l} \ m_j^2 m_k^2 
 \left[ \log(m_j^2) + \log(m_k^2) \right]^2  \nonumber \\ 
 &&{\kern-20pt}+ \Qol{l}{j} \ m_l^2 m_k^2 
 \left[ \log(m_l^2) + \log(m_k^2) \right]^2  \nonumber \\ 
 &&{\kern-20pt}+ \Qol{l}{k} \ m_l^2 m_j^2 
 \left[ \log(m_l^2) + \log(m_j^2) \right]^2. 
\label{mde2}
\end{eqnarray} 
The {\em special} points are $p^2=0, \infty$ and the roots of 
\begin{eqnarray} 
&&{\kern-20pt} \D  = \left[p^2+(m_1+m_2+m_3)^2\right] \nonumber \\ 
&&{\kern-20pt} \left[p^2+(m_1+m_2-m_3)^2\right] 
 \left[p^2+(m_1-m_2+m_3)^2\right] \nonumber \\ 
&&{\kern-20pt} \left[p^2+(m_1-m_2-m_3)^2\right]  = 0 \ , 
\end{eqnarray}
and $\Pol{l}{j}$ and $\Qol{l}{j}$ are polynomials in $p^2$
and in the masses, whose explicit expressions can be found in \cite{CCLR1}.

From these equations the analytic expressions for their first order
expansion were completed around the {\em special} points 
\cite{CCLR1,CCR1,CCR2,CCR3}: 
$p^2=0$;  $p^2=\infty$;  $p^2=-(m_1+m_2+m_3)^2$, 
the threshold; $p^2=-(m_1+m_2-m_3)^2$, the pseudo-thresholds.

To obtain numerical results for arbitrary values of $p^2$, a 
4th-order Runge-Kutta formula is implemented in a FORTRAN code, 
with a solution advancing path starting from the {\em special} points, 
so that also the first term in the expansion is necessary.

\begin{figure}[htb]
{\scalebox{.5}[.5]{\includegraphics*[40,20][460,250]{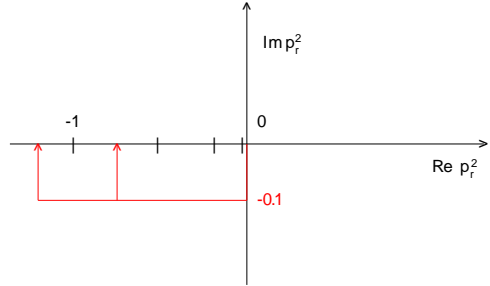}}}
\caption{Paths followed in the complex $p_r^2$ plane. On the real axis 
are indicated the positions of the threshold (-1) and pseudo-thresholds.}
\label{fig:cpxp2}
\end{figure}

The path followed starts usually from $p^2=0$ and moves in the lower 
half complex plane of $p_r^2 \equiv p^2/(m_1+m_2+m_3)^2$, as shown 
in Fig.\ref{fig:cpxp2}, to avoid proximity to the other {\em special} 
points, which can cause loss in precision.
Values between {\em special} points can be safely reached 
through a complex path as also shown in Fig.\ref{fig:cpxp2}. 
For values of $p^2$ very close to a {\em special} point, we start from 
the analytical expansion at that {\em special} point. 
Subtracted differential equations are used when starting from $p^2=\infty$ 
or from threshold, as that points are not regular points of the 
MDE \Eq{mde1},\Eq{mde2}.

Remarkable self-consistency checks are easily provided by comparing 
the results obtained either starting from the same point and choosing 
different paths to arrive to the same final point, or choosing directly 
different starting points and again arriving to the same final point.

The execution of the program is rather fast and precise: with an 
Intel Pentium III of 1 GHz we get values with 7 digits requiring 
times ranging from a fraction of a second to 10 seconds of CPU, 
and with 11 digits from few tens of seconds to one hour. 

If $\Delta=L/N$ is the length of one step, $L$ is the length of the 
whole path and $N$ the total number of steps, the 4th-order 
Runge-Kutta formula discards terms of order $\Delta^5$, so the 
whole error behaves as $\epsilon_{RK} = N \Delta^5 = L^5/N^4$, and a proper 
choice of $L$ and $N$ allows the control of the precision. 

Indeed we estimate the relative error, as usual, by comparing a value 
obtained with $N$ steps with the one obtained with $N/10$ steps, 
$\epsilon_{RK} = [V(N)-V(N/10)]/V(N)$, to which we add a cumulative 
rounding error $\epsilon_{cre} = \sqrt{N} \times 10^{-15}$, 
due to our 15 digits double precision FORTRAN implementation. 

The general massive sunrise graph is numerically well studied in literature 
and several numerical methods are developed, such as multiple expansions 
\cite{BBBS}, or numerical integration \cite{BBBS,BBBB,GvdB,PT,GKP,P}.
Comparisons are presented in \cite{CCR3} with some values available in the 
literature \cite{BBBS,P} with excellent agreement (up to more 
than 11 digits).

\section{Perspectives}

The presented method for numerically advancing the solutions of the MDE 
is rather precise and competitive with other available methods 
for numerical MI calculations.

Rather than conclusions it is more appropriate at this stage to present
perspectives. It seems to be possible to complete the 2-loop self-mass 
for arbitrary internal masses and we have almost completed the 4-denominators
case \cite{CCGR}. 

We think that the extension to graphs with more loops or legs do not 
present serious problems, even if the growth in the number of MI increases 
the computing time. 

It is worth to mention that the method relies on the same MDE, which are 
used also for analytic calculations, so it provides a 'low-cost' comforting 
cross-check for those results. 
\def\NP{{\sl Nucl. Phys.}} 
\def\PL{{\sl Phys. Lett.}} 
\def\PR{{\sl Phys. Rev.}} 
\def\PRL{{\sl Phys. Rev. Lett.}} 
\def\NC{{\sl Nuovo Cim.}}
\def\APP{{\sl Acta Phys. Pol.}}
\def\ZP{{\sl Z. Phys.}}
\def\MPL{{\sl Mod. Phys. Lett.}} 
\def\EPJ{{\sl Eur. Phys. J.}} 
\def\IJMP{{\sl Int. J. Mod. Phys.}}

\end{document}